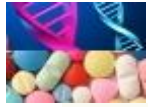

# On the Mass and Magnetic Field of the Neutron Star in the Ultraluminous X-Ray Source NGC 300 ULX1

Mehmet Hakan Erkut[1*]

[1*] Istanbul Bilgi University, Faculty of Engineering and Natural Sciences, 34060, İstanbul, Turkey, (ORCID: 0000-0003-1054-264X), mherkut@gmail.com

## Abstract

The accreting compact objects in most of ultraluminous X-ray sources (ULXs) are likely to be neutron stars rather than black holes as suggested by the recent detection of periodic pulsations from some of these sources located in neighboring galaxies and one ULX that has hitherto been discovered in our own galaxy. As a member of the ULX family, NGC 300 ULX1 is a new pulsating ULX (PULX) spinning up at substantially high rates compared with other PULXs. In this paper, the strength of the magnetic field on the surface of the neutron star is inferred from the energy of the cyclotron absorption line detected in the pulsed X-ray spectrum of NGC 300 ULX1 and the plausible ranges for the neutron-star mass and beaming fraction are estimated using the observed spin period and period derivative of the pulsar and the measured X-ray flux of the source. Our analysis favors proton cyclotron resonance scattering as a viable mechanism to account for both the observed cyclotron energy and high spin-up rates provided that the absorption line is generated close to the surface of the neutron star.

# Aşırı Parlak X-Işın Kaynağı NGC 300 ULX1'deki Nötron Yıldızının Kütlesi ve Manyetik Alanı Üzerine

## Öz

Şu ana kadar galaksimizde keşfedilen bir aşırı parlak X-ışın kaynağı (APX) ve komşu galaksilerde bulunan bazı APX'lerden geldiği kısa süre önce saptanan periyodik pulsasyonların düşündürdüğü üzere, APX'lerin çoğunda kütle yığıştıran yoğun cisimler karadeliklerden daha çok, büyük olasılıkla, nötron yıldızlarıdır. APX ailesinin bir üyesi olan NGC 300 ULX1, diğer pulsasyonlu APX'ler (PAPX) ile karşılaştırıldığında oldukça yüksek oranlarda dönüşü hızlanan yeni bir PAPX'tir. Bu makalede, NGC 300 ULX1'in pulsasyonlu X-ışın tayfında saptanan siklotron soğurma çizgisinin enerjisinden nötron yıldızının yüzeyindeki manyetik alan yeğinliği çıkarsanmakta ve kaynağın X-ışın akısı ile pulsarın gözlenen dönme periyodu ve periyot türevi kullanılarak nötron yıldızının kütlesi ve hüzmeleme oranı için olası aralıklar kestirilmektedir. Analizimiz, soğurma çizgisinin nötron yıldızı yüzeyine yakın üretilmesi koşuluyla, gözlenen siklotron enerjisi ve dönme hızındaki yüksek artış oranlarının her ikisini de açıklayacak yeterli bir mekanizma olması bakımından proton siklotron rezonans saçılımını desteklemektedir.

**Anahtar Kelimeler:** Nötron yıldızları, Yığışma, X-ışın çift yıldızları.

---

[*] Corresponding Author: mherkut@gmail.com

# 1. Introduction

Among the most energetic sources such as active galactic nuclei (AGN), gamma-ray bursts, and supernovae are the ultraluminous X-ray sources (ULXs) observed in nearby galaxies with typical X-ray luminosities in the ~ $10^{39}-10^{41}$ erg s$^{-1}$ range. ULXs can be distinguished from AGN based on their off-nuclear positions in the host galaxy. The Eddington limit for the luminosity of a stellar mass object of ~ $1-10$ M$_\odot$ is exceeded by the X-ray luminosity of a ULX under the assumption of isotropic emission. A few members of this family, which are also known as hyperluminous X-ray sources, observed in the higher luminosity tail of the ULX population with luminosities ≥ $10^{41}$ erg s$^{-1}$ are still believed to be potentially powered by the intermediate mass black holes of $10^2-10^5$ M$_\odot$ accreting matter and emitting X-rays isotropically. It is however likely for most of the ULXs with the standard $10^{39}-10^{41}$ erg s$^{-1}$ luminosity range that the neutron stars and stellar mass black holes in high-mass X-ray binaries accrete matter at super-Eddington rates and emit radiation anisotropically with a geometrical beaming factor that accounts for the observed super-Eddington luminosities (Kaaret et al., 2017).

Coherent pulsations with periods in the $0.417-31.6$ s range have recently been detected in X-rays from seven ULXs of which one, namely, Swift J0243.6+6124 is galactic (Wilson-Hodge et al., 2018) and all the others, namely, M82 X-2, ULX NGC 7793 P13, ULX NGC 5907, NGC 300 ULX1, NGC 1313 X-2, and M51 ULX-7 are extragalactic (Bachetti et al., 2014; Fürst et al., 2016; Israel et al., 2017; Carpano et al., 2018; Sathyaprakash et al., 2019; Rodríguez Castillo et al., 2020). The discovery of periodic signals from pulsating ULXs (PULXs) has firmly confirmed that the majority of accreting compact stars in ULXs are indeed stellar-mass objects and presumably neutron stars being outnumbered compared to black holes. The observed X-ray luminosities of PULXs above the Eddington limit have been explained either by the effect of moderate beaming (Bachetti et al., 2014) or by the presence of sufficiently high neutron-star magnetic fields to reduce the scattering cross section and therefore increase the level of maximum critical luminosity (Ekşi et al., 2015). The recent analysis of PULXs has revealed that both the beaming of X-ray emission and the reduction of cross section due to the magnetic field of the neutron star should be taken into account for studying the feasibility of different spin and luminosity states (Erkut et al., 2020).

The lack of pulsations should not be considered as a direct evidence for black holes in non-pulsating ULXs. An optically thick environment enveloping the neutron-star magnetosphere and hence wiping off the pulsations imprinted in the X-ray emission (Ekşi et al., 2015), a super-Eddington propeller stage where the spindown power brightens the source even in the absence of accretion (Erkut et al., 2019) or an emission with a small fraction of beaming (Erkut et al., 2020) can be the reason of why we do not observe pulsations from other ULXs. The presence of a cyclotron resonance scattering feature (CRSF) in the spectrum of a ULX is a signature of the neutron-star magnetic field on the X-ray emission whether or not pulsations are detected. CRSFs have been observed in the X-ray spectra of the PULX NGC 300 ULX1 (Walton et al., 2018) and the non-pulsating ULX M51 ULX-8 (Brightman et al., 2018) providing further support in favor of neutron stars rather than black holes for the identification of the compact object in ULXs.

In this paper, the neutron star's surface magnetic field is inferred from the energy of a potential cyclotron absorption line detected at ~ 13 keV in the pulsed spectrum of NGC 300 ULX1. Using the magnetic field estimates in line with both the electron and proton cyclotron interpretations of the observed feature, the allowed region is determined in the parameter space of beaming fraction and neutron-star mass. In addition to the range estimate for the mass and beaming fraction, the state of the accretion disk around the neutron star is discussed taking into account the relative positions of the spherization and corotation radii with respect to the innermost disk radius at sufficiently high rates of mass transfer and the structure of the magnetic-field-threaded inner disk boundary region is revealed.

# 2. Material and Method

## 2.1. Assumptions and Equations

The suppression of the electron scattering opacity by the high-strength magnetic field on the stellar surface leads to an increase in the maximum critical luminosity $L_c$ of an accreting neutron star (Paczynski, 1992). For negligible field strengths, the Eddington limit defines the maximum critical luminosity, that is, $L_c = L_E$. The maximum critical luminosity of an accreting neutron star can be written as

$$L_c = \left[1 + 311\left(\frac{B}{B_c}\right)^{4/3}\right]L_E \quad (1)$$

where the magnetic field on the neutron star surface is $B$, the value of the quantum critical field for electrons is

$$B_c = \frac{m_e^2 c^3}{\hbar e} \cong 4.4 \times 10^{13} \text{ G}, \quad (2)$$

and the Eddington luminosity is

$$L_E = 4\pi G M_* m_p c / \sigma_T \approx 1.8 \times 10^{38} M_{1.4} \text{ erg s}^{-1} \quad (3)$$

(Erkut et al., 2020). In Equations (2) and (3), $M_{1.4}$ represents the neutron-star mass $M_*$ in units of 1.4 solar mass, $\sigma_T$ stands for the Thomson cross section of the electron, and $e$ is the elementary charge. The electron and proton masses are given by $m_e$ and $m_p$, respectively. The fundamental constants such as the reduced Planck's constant, the speed of light, and the gravitational constant are denoted by their usual symbols.

The relation between the luminosity and flux of the source in X-rays can be expressed in terms of the source distance $d$ and the beaming fraction $b \leq 1$ as

$$L_X = 4\pi b d^2 F_X, \quad (4)$$

where $L_X$ is the intrinsic X-ray luminosity of the source and $F_X$ is the unabsorbed source flux in X-rays measured by a distant observer. Assuming that the gravitational potential energy is mostly released in X-rays,

$$L_X = G M_* \dot{M}_* / R_* = \varepsilon \dot{M}_* c^2 \quad (5)$$

yields the X-ray luminosity in terms of the neutron star's mass and radius $R_*$, the efficiency of gravitational energy release $\varepsilon$, and the mass-accretion rate $\dot{M}_*$ onto the neutron-star surface. In a steady state, the mass accretion rate is the same as the mass-inflow rate at the innermost radius $R_{in}$ of the accretion disk interacting with the neutron star. For ULXs, where the donor star in a high-mass X-ray binary presumably transfers mass at super-Eddington (supercritical) rates,

$$\dot{M}_0 > \dot{M}_E = \frac{L_E}{\varepsilon c^2}, \qquad (6)$$

to the disk around the neutron star, the spherization of the accretion flow due to high radiation pressure inside the so-called spherization radius $R_{sp}$ occurs if $R_{in} < R_{sp}$. The spherization radius for a supercritical accretion disk around a neutron star of mass $M_*$ and radius $R_*$ is given by

$$R_{sp} = \frac{27 \sigma_T GM_* \dot{M}_0}{8\pi m_p R_* c^3} \qquad (7)$$

(Shakura & Sunyaev, 1973). The self regulation of the mass inflow in the disk such that the Eddington limit can be slightly exceeded by the local disk luminosity leads to an outflow of matter for $R_{in} < R < R_{sp}$. The mass accretion rate onto the neutron-star surface is then determined only by a fraction of the mass transfer rate, i.e., $\dot{M}_* = (R_{in}/R_{sp}) \dot{M}_0$. The spherization of the disk is not realized even at high mass-accretion rates provided $R_{in} > R_{sp}$. In this regime, the mass-transfer rate directly determines the mass-accretion rate, that is, $\dot{M}_* = \dot{M}_0$ (Shakura & Sunyaev, 1973; Erkut et al., 2020).

The inner radius of a supercritical accretion disk truncated by the magnetosphere of a neutron star can be derived from the conservation of angular momentum, which can be written in a simple form as

$$\frac{d}{dR}(\dot{M} R^2 \Omega) = -R^2 B_\phi^+ B_z \qquad (8)$$

at a disk radius $R$ inside the boundary layer within the neighborhood of $R_{in}$. Equation (8) reflects the close balance between the magnetic and material stresses in this innermost disk region. The angular velocity $\Omega$ of the boundary region matter deviates from its Keplerian value $\Omega_K(R_{in})$ to match with the rotation frequency of the neutron star under the action of magnetic stresses, which are due to coupling between the toroidal magnetic field component at the disk surface $B_\phi^+$ and the poloidal magnetic field component $B_z$ of magnetospheric origin (Ghosh & Lamb, 1979). Integrating Equation (8) over the radial width of the boundary region, $\delta R_{in}$, while taking into account the supercritical disk regimes with $R_{in} < R_{sp}$ and $R_{in} > R_{sp}$, Erkut et al. (2020) found

$$R_{in} = \left( \frac{B^2 R_*^5 \sqrt{GM_*} \delta}{16\pi b d^2 F_X} \right)^{2/7} \qquad (9)$$

for the inner radius of the accretion disk assuming that the poloidal field has a dipolar nature giving rise to the toroidal field

$$\frac{B_\phi^+}{B_z} = \frac{\Omega_* - \Omega_K(R_{in})}{\Omega_K(R_{in})} = \omega_* - 1 \qquad (10)$$

as a result of the differential rotation between the innermost disk matter and the neutron-star magnetosphere. In Equation (10), the fastness parameter,

$$\omega_* \equiv \frac{\Omega_*}{\Omega_K(R_{in})} = \left( \frac{R_{in}}{R_{co}} \right)^{3/2}, \qquad (11)$$

is defined in terms of the ratio of the inner radius to the corotation radius, $R_{co}$, which is deduced from $\Omega_K(R_{co}) = \Omega_*$ as

$$R_{co} = \left( \frac{GM_*}{\Omega_*^2} \right)^{1/3} = \left( \frac{GM_* P^2}{4\pi^2} \right)^{1/3}, \qquad (12)$$

where $P$ is the spin period of the neutron star. Note that the rotation frequency $\Omega_*$ of the neutron star should not exceed $\Omega_K(R_{in})$ in order for the matter in the boundary region to overcome the centrifugal barrier and be accreted by the neutron star. Therefore, $\omega_* < 1$ ($R_{in} < R_{co}$) must be satisfied in an accreting system.

The mass accreting objects are expected to spin up as they acquire the excess of angular momentum carried by the infalling matter. The accretion torque acting on a neutron star of moment of inertia $I$ can therefore be expressed as

$$N = -2\pi I \dot{P}/P^2 = \dot{M}_* \sqrt{GM_* R_{in}}. \qquad (13)$$

Using Equations (4), (5), (9), (11), and (12), it follows from Equation (13) that

$$B = \frac{\omega_*}{R_*^3} \sqrt{\frac{2GM_* I |\dot{P}|}{\pi \delta}} \qquad (14)$$

for the polar surface-field strength and

$$\omega_* = \frac{\pi GM_*}{4} \left( \frac{I |\dot{P}|}{bd^2 F_X P^{7/3} R_*} \right)^3 \qquad (15)$$

for the fastness parameter (Erkut et al., 2020). Next, Equations (14) and (15) are combined to solve for the beaming fraction

$$b = \frac{\pi^{1/6} \sqrt{GM_*} (I|\dot{P}|)^{7/6}}{\sqrt{2} \delta^{1/6} B^{1/3} R_*^2 d^2 F_X P^{7/3}} \qquad (16)$$

in terms of the measurable parameters such as $d$, $F_X$, $P$, and $\dot{P}$ in addition to $B$, $M_*$, $R_*$, $I$, and $\delta$ inferred from model dependent interpretations of observations. Note from Equation (15) that the lower limit for the beaming fraction corresponds to the upper limit of the fastness parameter, that is, for the lower limit of $b$,

$$b_{min} = \left( \frac{\pi GM_*}{4} \right)^{1/3} \frac{I|\dot{P}|}{d^2 F_X P^{7/3} R_*} \qquad (17)$$

is obtained by setting $\omega_* = 1$ in Equation (15).

The X-ray luminosity of an accreting neutron star cannot exceed the maximum critical luminosity (Equation 1). The subcritical luminosity condition, $L_X \leq L_c$, is considered to estimate the upper limit of the beaming fraction. Using Equations (1) and (4), $L_X \leq L_c$ yields

$$b \leq \left[1 + 311\left(\frac{B}{B_c}\right)^{4/3}\right] \frac{L_E}{4\pi d^2 F_X}. \quad (18)$$

The upper limit of the beaming fraction is given by $b = 1$ unless the right hand side of the inequality (Equation 18) is less than 1. The upper limit of $b$ is therefore expressed as

$$b_{max} = \min\left\{1, \left[1 + 311\left(\frac{B}{B_c}\right)^{4/3}\right]\frac{L_E}{4\pi d^2 F_X}\right\}. \quad (19)$$

The lower and upper limits given in Equations (17) and (19) can be useful in exploring the physically plausible region in the phase space defined by the parameters in Equation (16).

## 2.2. Cyclotron Feature and Magnetic Field

The resonant scattering of photons by electrons and protons moving in a magnetic field with quantized cyclotron orbits of energies known as Landau levels may lead to the formation of absorption features in the X-ray spectrum of a magnetized neutron star. The magnetic poles of the neutron star are therefore seen as the prime candidates for the regions where these absorption lines can be generated following the interaction of the radiation emitted by the neutron star with the charged particles of the infalling plasma along the field lines.

The energy of each cyclotron line is an integer multiple of $\hbar\omega_{cyc}$, where $\omega_{cyc} = eB/(mc)$ is the cyclotron frequency of a particle of charge $e$ and mass $m$. The observed energy of the cyclotron feature is expected to be redshifted with respect to the genuine line energy due to the strong gravitational field of the neutron star. The energy of the potential cyclotron feature to be detected in the source spectrum is then

$$E_{cyc} = \frac{n\hbar e}{(1+z)mc} B. \quad (20)$$

The gravitational redshift in Equation (20) is

$$z = \frac{1}{\sqrt{1 - \frac{2GM_*}{c^2 R_{cyc}}}} - 1, \quad (21)$$

where $R_{cyc}$ is the line formation radius. The quantum number $n = 1$ yields the fundamental line corresponding to the transition from the ground state to the first excited state and the quantum number $n = 2$ gives the first harmonic of the fundamental line resulting from the transition from the ground state to the second excited state (Staubert et al., 2019).

Substituting the electron mass $m_e$ for the mass of the charged particle $m$ in Equation (20), the magnetic field in the medium, where the electron cyclotron resonance scattering feature (eCRSF) is expected to form, is estimated as

$$B_e \cong \frac{E_{cyc}}{11.6\,\text{keV}} \left(\frac{1+z}{n}\right) \times 10^{12}\,\text{G}. \quad (22)$$

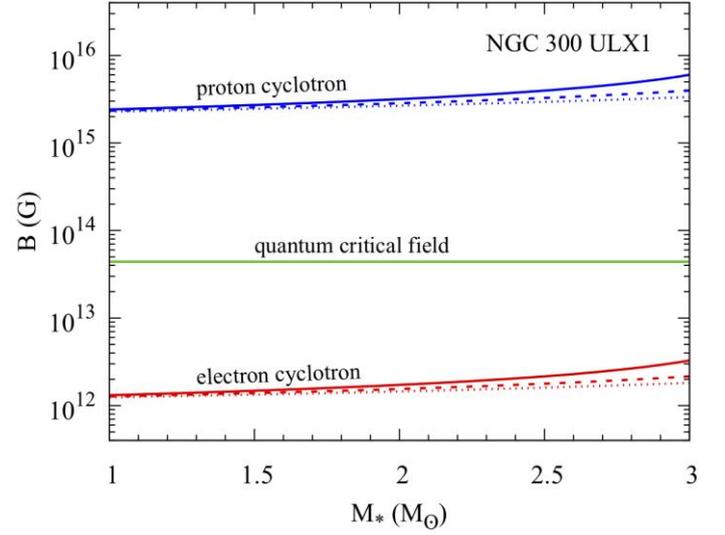

*Figure 1. Magnetic field estimate for a wide range of the neutron-star mass using the CRSF detected at 12.8 keV in the X-ray pulsed spectrum of NGC 300 ULX1 with regard to electrons and protons. The solid, dashed, and dotted curves are obtained for the neutron-star radii of 10, 12, and 14 km, respectively.*

Similarly, the proton mass $m_p$ is substituted for the particle mass $m$ in Equation (20) to estimate the magnetic field as

$$B_p \cong \frac{E_{cyc}}{6.32\,\text{keV}} \left(\frac{1+z}{n}\right) \times 10^{15}\,\text{G} \quad (23)$$

for the medium where the proton cyclotron resonance scattering feature (pCRSF) would be produced. Note that the $B$ values estimated according to the eCRSF and pCRSF interpretations of the observed cyclotron line energy differ from one another by orders of magnitude.

## 3. Results and Discussion

### 3.1. Magnetic Field of the Neutron Star in NGC 300 ULX1

A direct method for the estimation of the magnetic field strength on the surface of an accreting neutron star is to measure the centroid energy $E_{cyc}$ of a CRSF detected in the X-ray spectrum of the source. Based on the phase-resolved broadband spectroscopy, a potential CRSF at $E_{cyc} \approx 12.8$ keV was detected in the pulsed spectrum of NGC 300 ULX1 using XMM-Newton and NuSTAR data (Walton et al., 2018). The same data set had also revealed the presence of coherent pulsations with a neutron-star spin period of $\sim 31.6$ s (Carpano et al., 2018).

The line energy at 12.8 keV is substituted for $E_{cyc}$ in Equations (22) and (23) to estimate the surface magnetic field of the neutron star based on both the electron and proton interpretations of the observed CRSF, as shown in Figure 1. The absorption feature detected in the pulsed spectrum is presumed to correspond to the fundamental cyclotron line given by $n = 1$, as also suggested by Walton et al. (2018). Setting $R_{cyc} = R_*$ in Equation (21), the redshift factor in Equations (22) and (23) is calculated in line with the assumption that the cyclotron line is produced near the surface of the neutron star at an emission height that is negligible compared to the neutron-star radius as is likely the case with some, if not all, of the sources accreting at supercritical rates (Becker et al., 2012). A brief speculation is presented in Section 4 on the possible consequences of a higher altitude being well above the neutron-star surface as a choice for the emission region where the CRSF could be generated.

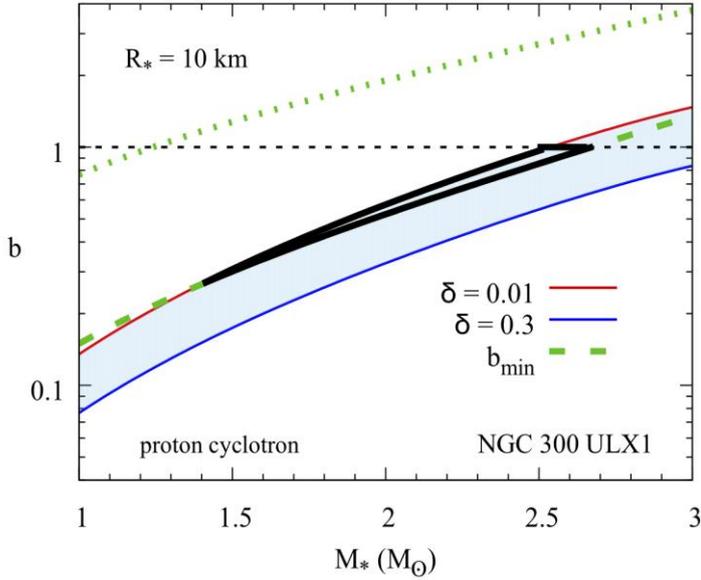

*Figure 2. Beaming fraction as a function of the mass of the neutron star in NGC 300 ULX1. The shaded area bounded by the solid black curve represents the allowed region for the mass and beaming fraction according to the proton cyclotron interpretation when the neutron-star radius is 10 km. See the text for the detailed explanation of all curves in the plot.*

The variation of the surface magnetic field over a wide range of neutron-star mass is displayed in Figure 1. The 1−3 $M_\odot$ range for the mass is also in agreement with the range of neutron-star masses measured from pulsar timing (Lattimer, 2019). In Figure 1, the curves of magnetic field estimate according to both electron and proton cases are obtained for three different neutron-star radii in the 10−14 km range that covers almost all of the equations of state supporting the existence of neutron stars of at least 2 $M_\odot$ in line with the recent discovery of the massive pulsar J0740+6620 (Cromartie et al., 2020). The ranges under consideration for the neutron-star mass and radius are also consistent with the allowed region determined in the mass-radius space via generalization of the equations of state to higher densities (Lattimer, 2019).

Note from Figure 1 that the magnetic field increases with the mass for a given value of the radius since the redshift factor in Equations (22) and (23) becomes more pronounced for neutron stars of higher compactness. The field values estimated by both the eCRSF and pCRSF interpretations are higher for smaller radii at a given mass for the same reason. The magnetic field inferred from electrons is lower by about three orders of magnitude than the field value inferred from protons, which indeed exceeds the quantum critical magnetic field for electrons (Equation 2). The magnetic field inferred from eCRSF approaches its quantum critical value only when the electron's cyclotron and rest-mass energies are comparable.

The implications of the eCRSF and pCRSF interpretations on the neutron-star mass are studied in the next subsection.

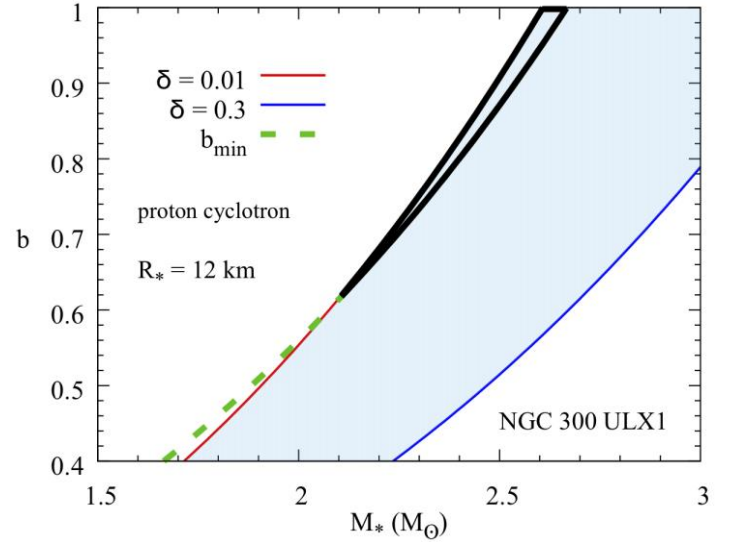

*Figure 3. Beaming fraction as a function of the mass of the neutron star in NGC 300 ULX1. The shaded area bounded by the solid black curve represents the allowed region for the mass and beaming fraction according to the proton cyclotron interpretation when the neutron-star radius is 12 km.*

### 3.2. Mass of the Neutron Star in NGC 300 ULX1

The combined 2016 observations of the PULX NGC 300 ULX1 by XMM-Newton and NuSTAR have revealed the presence of periodic pulsations in the X-ray data with an average spin period and period derivative being measured as $P = 31.6$ s and $\dot{P} = -5.56 \times 10^{-7}$ s s$^{-1}$, respectively. The 0.3−30 keV X-ray flux of the PULX has been obtained as $F_X \approx 1.11 \times 10^{-11}$ erg s$^{-1}$ cm$^{-2}$, which is equivalent to an isotropic X-ray luminosity of $L_X \approx 4.7 \times 10^{39}$ erg s$^{-1}$ for a source distance of $d \approx 1.88$ Mpc (Carpano et al., 2018). The isolated spectrum of the pulsed component in the same 2016 X-ray data of NGC 300 ULX1 has subsequently disclosed a potential cyclotron absorption line feature at $E_{cyc} \approx 12.8$ keV (Walton et al., 2018).

In the following analysis, the above-mentioned numerical values are substituted for all the measured parameters such $P$, $\dot{P}$, $F_X$, and $d$ in Equation (16) and $E_{cyc}$ in Equations (22) and (23) from which $B$ in Equation (16) can be inferred according to both the eCRSF and pCRSF interpretations. In Equation (16), the relative width of the boundary region is allowed to vary between $\delta \approx 0.01$ and $\delta \approx 0.3$ (Erkut & Çatmabacak, 2017; Erkut et al., 2020) and the empirical formula with dependence on both mass and radius is used for the neutron-star moment of inertia (Lattimer & Schutz, 2005; Erkut et al., 2020). In the same equation, the neutron-star mass is also allowed to vary in the 1−3 $M_\odot$ range for the three representative values of the neutron-star radius in the 10−14 km range.

No solution is found for the beaming fraction as a function of mass in the case of eCRSF within the given ranges of mass and radius and even beyond the limits of these ranges as the $b$ values computed with the help of Equation (16) do not satisfy $b < b_{max}$ (Equation 19) and, moreover, the condition that $b_{max} > b_{min}$ is violated. It is not difficult to note from Equations (17) and (19) that the lower and upper constraints on the beaming fraction may start overlapping for sufficiently high values of the period derivative as observed in the case of NGC 300 ULX1 and low values of the surface magnetic field as estimated in line with the eCRSF interpretation (Figure 1).

According to the pCRSF interpretation of the observed line energy, however, physically plausible solutions are obtained in the plane of beaming fraction versus mass as shown in Figures 2 and 3. For a neutron-star radius of 10 km, the variation of the beaming fraction with the mass of the neutron star is shown in Figure 2. The possible values of the beaming fraction in the absence of the upper and lower limits of $b$ lie within the shaded area bounded by the solid red and blue curves labelled with $\delta = 0.01$ and $\delta = 0.3$, respectively. The inclusion of $b_{min}$ shown by the dashed green curve and of $b_{max} = 1$ given by the horizontal dashed black line severely constrains the possible $b$ and $M_*$ values. The solid black curves in Figures 2 and 3 determine the boundaries of the allowed regions in the $b$ versus $M_*$ plane with all possible constraints being taken into account. The $b$ values above $b_{min}$ guarantee the slow rotator condition for accretion to occur, that is, $\omega_* < 1$ assuring that $R_{in} < R_{co}$. The dotted green curve passing well above the shaded area corresponds to the upmost limit of the beaming fraction for the inner disk radius to lie outside the spherization radius, i.e.,

$$b < \frac{c}{3d^2 F_X} \left(\frac{GM_* m_p}{2\sigma_T}\right)^{1/3} \left(\frac{I|\dot{P}|}{R_* P^2}\right)^{2/3} \quad (24)$$

for $R_{in} > R_{sp}$. Note from Figure 2 that all values of the beaming fraction and neutron-star mass enclosed by the allowed region, which is encircled by the solid black curve, are below the upmost limit in Equation (24) and thus are valid for a system where the mass transfer rates might be super-Eddington and yet the disk remains geometrically thin. The same condition is also satisfied in Figure 3 obtained for a neutron star of radius 12 km. No solution is found for the beaming fraction as a function of mass for neutron-star radii $\geq 14$ km even in the case of pCRSF as $b_{min}$ passes above the upper boundary with $\delta = 0.01$ without crossing the shaded area.

Under the assumption that the absorption line is generated near the surface of the neutron star, i.e., $R_{cyc} \approx R_*$, the ranges deduced from Figures 2 and 3 for the neutron-star mass and beaming fraction, based on the pCRSF interpretation of the observed feature, are $1.40\ M_\odot < M_* < 2.66\ M_\odot$ and $0.26 < b < 1$ for $R_* = 10$ km and $2.09\ M_\odot < M_* < 2.66\ M_\odot$ and $0.61 < b < 1$ for $R_* = 12$ km, respectively, on the contrary of the null result based on the eCRSF interpretation of the same feature.

### 3.3. Boundary Region of the Disk around the Neutron Star in NGC 300 ULX1

In the innermost regions of the accretion around a strongly magnetized neutron star, the disk matter rotating in nearly Keplerian orbits couples with the neutron-star magnetosphere via Rayleigh-Taylor and Kelvin-Helmholtz instabilities. The electric currents induced within the neighborhood of the inner disk radius $R_{in}$ screen most of the dipolar magnetic field of the neutron star over the radial extent $\delta R_{in}$ of the so-called boundary region, also known as the transition zone or boundary layer (Ghosh & Lamb, 1979). The poloidal component of the magnetic field is expected to be attenuated along a screening length of radial width $\sim \delta R_{in}$. The steep radial dependence of the poloidal field component resulting from this efficient attenuation provides the sufficiently large magnetic stresses required to maintain the angular momentum balance against the material stresses in the boundary region (Equation 8).

As can be noted from Equation (8), the change of angular momentum flux carried by the inner disk matter, also being known as material stresses, can be realized only if the matter's rotation rate $\Omega$ in the boundary region deviates from its Keplerian value and finally matches to the spin frequency of the neutron star with $\Omega_* < \Omega < \Omega_K(R_{in})$. In the inner region of an accretion disk such as the one around the neutron star in the PULX NGC 300 ULX1, a sub-Keplerian boundary region is therefore expected to be formed. The $z$ component of the strong magnetic field in the boundary region can be deduced from the poloidal component of the induction equation,

$$\frac{\partial B_R}{\partial z} - \frac{\partial B_z}{\partial R} \approx -\frac{v_R B_z}{\eta}, \quad (25)$$

where $v_R$ is the radial drift velocity of the matter in the disk, $B_R$ is the radial field component, and $\eta$ is the magnetic diffusivity (Erkut & Alpar, 2004; Erkut, 2005). Assuming that the $z$ dependence of $B_z$ is negligible whereas $B_R$ obeys the field antisymmetry across the midplane of the disk,

$$B_R(R, z) = \left(\frac{z}{H}\right) B_R^+(R), \quad (26)$$

the simplified version of Equation (25) can be written as

$$\frac{dB_z}{dR} \approx -\frac{B_z}{\alpha_D \delta R_{in}}, \quad (27)$$

noting that $|B_r / B_z| \ll 1$ at the disk surface ($z = H$) and the magnetic diffusivity in the boundary region can be prescribed as

$$\eta = -\alpha_D v_R \delta R_{in}, \quad (28)$$

where the coefficient of magnetic diffusivity is given by a dimensionless constant $\alpha_D \leq 1$ (Erkut, 2005). The solution of Equation (27) is expressed in a dimensionless form as

$$b_z(x) = b_0 \exp\left(-\frac{x-1}{\alpha_D \delta}\right) \quad (29)$$

using $x \equiv R / R_{in}$, $b_z \equiv B_z / B_t$, and $b_0 \equiv B_z(R_{in}) / B_t$, with

$$B_t = -\frac{\mu}{R_{in}^3} = -\frac{BR_*^3}{2R_{in}^3} \quad (30)$$

being the typical value of the magnetic field in the boundary region, which can be estimated by the dipolar field strength at the inner disk radius. Here, $\mu$ is the magnetic dipole moment of the neutron star in terms of the polar surface-field strength $B$. Applying coordinate stretching in the boundary region, the field solution in Equation (29) can be written as

$$b_z(X) = b_0 \exp\left(-X/\alpha_D\right) \quad (31)$$

with the help of a new coordinate $X \equiv (x-1) / \delta$ (Erkut, 2005).

Next, Equation (8) is solved for the radial profile of the sub-Keplerian angular velocity $\Omega$ in the boundary region. First, Equation (10) is substituted into Equation (8) to express the toroidal field component at the disk surface in terms of $B_z$. Then, $\omega \equiv \Omega/\Omega_K(R_{in})$ is defined in Equation (8) as a non-dimensional parameter in addition to $x$ and $b_z$ to obtain

$$\frac{\dot{M}_* \sqrt{GM_*}}{B_t^2 R_{in}^{5/2} \delta} \frac{d\omega}{dX} = -(\omega_* - 1)b_z^2 \quad (32)$$

using $\dot{M} = \dot{M}_0 = \dot{M}_*$ for $R_{in} > R_{sp}$, as suggested by the pCRSF interpretation of the observed cyclotron line in the pulsed spectrum of NGC 300 ULX1. Note that the coefficient in front of the first derivative of $\omega$ on the left-hand side of Equation (32) is of order unity. Choosing therefore

$$\frac{\dot{M}_* \sqrt{GM_*}}{B_t^2 R_{in}^{5/2} \delta} = 1 \quad (33)$$

simply yields the expression for $R_{in}$ in Equation (9) if Equations (4), (5), and (30) are employed. Substituting Equation (31) for $b_z$, it follows from Equation (32) and (33) that

$$\omega(X) = 1 - (1 - \omega_*) \exp\left(-2X/\alpha_D\right) \quad (34)$$

satisfying the inner boundary condition $\omega(0) = \omega_*$ (i.e., $\Omega = \Omega_*$) and the outer boundary condition $\omega(X \to \infty) = 1$, that is, $\Omega = \Omega_K(R_{in})$ for $b_0 = (2/\alpha_D)^{1/2}$.

## 4. Conclusions and Recommendations

Our analysis recommends that the proton cyclotron scattering of photons is more likely to account for the observed absorption feature in the pulsed spectrum of NGC 300 ULX1 than the cyclotron scattering of photons by electrons provided that the production region of the cyclotron line is close enough to the surface of the neutron star. The strong dipolar field of magnetar strength suggested by the pCRSF interpretation of the observed line energy causes the inner disk radius to attain sufficiently large values and hence raises the efficient lever arm to generate a sufficiently large torque needed to explain the observed spin-up rates ($\dot{P} \sim -10^{-7}$ s s$^{-1}$) as well.

In the case of eCRSF, the magnetic field inferred from the observed line energy is always much weaker than the field deduced from the same energy according to pCRSF. The only option for the magnetic field estimated by the presumption of eCRSF to reach sufficiently high values to account for the large spin-up rates measured from the combined 2016 XMM-Newton and NuSTAR data of NGC 300 ULX1 is to allow the line generating region to lie well above the surface of the neutron star. The neutron star is then expected to have a surface magnetic field strength of

$$B = 8 \times 10^{12} \left(\frac{R_{cyc}}{2R_*}\right)^3 \left(\frac{B_e}{10^{12} G}\right) G \quad (35)$$

for a dipolar field. The magnetic field of the region of radius $R_{cyc}$, where the cyclotron line is produced, can be calculated using Equation (22). For the surface magnetic field strength of the neutron star in NGC 300 ULX1, $B \sim 10^{13}$ G is inferred from Equation (35) if $R_{cyc} \sim 2R_*$. For surface magnetic fields as high as $10^{13}$ G, Erkut et al. (2020) had shown that a marginal solution satisfying the observed spin-up rates of the PULX NGC 300 ULX1 at subcritical luminosity can only be found when the beaming fraction is close to its maximum value $b = 1$.

## 5. Acknowledge